\documentclass[aps, reprint, pre, onecolumn, notitlepage]{revtex4-1}
\usepackage{dcolumn}%
\usepackage{epsfig, graphics, amssymb, amsmath, subeqnarray, color, bm}
\usepackage{epstopdf}

\def \lb{\left}
\def \rb{\right}

\def \d{\,\text{d}}

\def \bF{\mathbf{F}}

\def \bR{\mathbf{R}}

\def \bT{\mathbf{T}}
\def \bU{\mathbf{U}}

\def \bV{\mathbf{V}}

\def \bX{\mathbf{X}}

\def \be{\mathbf{e}}

\def \bf{\mathbf{f}}

\def \bt{\mathbf{t}}
\def \bu{\mathbf{u}}

\def \bx{\mathbf{x}}

\def \buh{\hat{\bu}}

\def \max{\text{max}}

\definecolor{dgreen}{rgb}{0.0, 0.4, 0.0}

\begin{document}

\title{Characteristics of undulatory locomotion in granular media}
\author{Zhiwei Peng}
\affiliation{Department of Mechanical Engineering, Institute of Applied Mathematics,\\
University of British Columbia, Vancouver, BC, V6T-1Z4 Canada}
\author{On Shun Pak\footnote{Electronic mail: opak@scu.edu}}
\affiliation{Department of Mechanical Engineering,
Santa Clara University,\\
Santa Clara, CA, 95053, USA}
\author{Gwynn J. Elfring\footnote{Electronic mail: gelfring@mech.ubc.ca}}
\affiliation{Department of Mechanical Engineering, Institute of Applied Mathematics,\\
University of British Columbia, Vancouver, BC, V6T-1Z4 Canada}
\date{\today}

\begin{abstract}
Undulatory locomotion is ubiquitous in nature and observed in different media, from the swimming of flagellated microorganisms in biological fluids, to the slithering of snakes on land, or the locomotion of sandfish lizards in sand. Despite the similarity in the undulating pattern, the swimming characteristics depend on the rheological properties of different media. Analysis of locomotion in granular materials is relatively less developed compared with fluids partially due to a lack of validated force models but recently a resistive force theory in granular media has been proposed and shown useful in studying the locomotion of a sand-swimming lizard. Here we employ the proposed model to investigate the swimming characteristics of a slender filament, of both finite and infinite length, undulating in a granular medium and compare the results with swimming in viscous fluids. In particular, we characterize the effects of drifting and pitching in terms of propulsion speed and efficiency for a finite sinusoidal swimmer. We also find that, similar to Lighthill's results using resistive force theory in viscous fluids, the sawtooth swimmer is the optimal waveform for propulsion speed at a given power consumption in granular media. The results complement our understanding of undulatory locomotion and provide insights into the effective design of locomotive systems in granular media.

\end{abstract}

%\keywords{undulatory locomotion; granular media; resistive force theory}

\maketitle

\section{Introduction}
\label{sec:intro}
Undulatory locomotion, the self-propulsion of an organism via the passage of  deformation waves along its body, is ubiquitous in nature \cite{gray1953undulatory,cohen2010swimming}. Flagellated microorganisms swim in fluids \cite{gray1955propulsion, chwang1971note, lighthill1976flagellar, keller1976swimming,purcell1977life, higdon1979hydrodynamic}, snakes slither on land  \cite{gray1946mechanism, guo2008limbless, Hu23062009,alben2013} and sandfish lizards (\textit{Scincus scincus}) undulate in granular substrates \cite{baumgartner2008investigating, maladen2009undulatory, ding2012mechanics}. Yet the underlying physics differ: from viscous forces  \cite{lauga2009hydrodynamics} in fluids to frictional forces \cite{maladen2009undulatory} in terrestrial media. The investigation of these undulatory mechanisms in different environments advances our understanding of various biological processes \cite{cohen2010swimming, fauci2006biofluidmechanics} and provides insights into the effective design of biomimetic robots \cite{williams2014self,maladen2011undulatory}. 

The swimming of microorganisms in Newtonian fluids, where viscous forces dominate inertial effects, is governed by the Stokes equations \cite{lauga2009hydrodynamics}. Despite the linearity of the governing equation, locomotion problems typically introduce geometric nonlinearity, making the problem less tractable \cite{sauzade11}. For slender bodies such as flagella and cilia, Gray and Hancock  \cite{gray1955propulsion} exploited their slenderness to develop a local drag model, called resistive force theory (RFT), which has been shown useful in modeling flagellar locomotion and the design of synthetic micro-swimmers \cite{lauga2009hydrodynamics,pak2014theoretical}. In this local theory, hydrodynamic interactions between different parts of the body are neglected and the viscous force acting on a part of the body depends only on the local velocity relative to the fluid. Using RFT, Lighthill showed that, for an undulating filament of infinite length, the sawtooth waveform is the optimal beating pattern maximizing hydrodynamic efficiency  \cite{lighthill1976flagellar}.

Locomotion in granular media is relatively less well understood due to their complex rheological features  \cite{zhang2014effective, goldman2014colloquium}. The frictional nature of the particles generates a yield stress, a threshold above which the grains flow in response to external forcing \cite{goldman2014colloquium}. Different from viscous fluids, the resistance experienced by a moving intruder originates from the inhomogeneous and anisotropic response of the granular force chains, which are narrow areas of strained grains surrounded by the unstrained bulk of medium \cite{albert1999slow}. At low locomotion speed, where the granular matter is in a quasi-static regime, the effect of inertia is negligible compared to frictional and gravitational forces from granular media \cite{ding2012mechanics}, which is similar to that of a low Reynolds-number fluid. In this regime, studies measuring the drag force of an intruder moving through a GM reveal that the drag force is independent of the speed of the intruder, but it increases with the depth of GM and proportional to the size of the intruder \cite{albert1999slow,hill2005scaling,schroter2007phase,zhou2007simul,seguin2011dense}.

Recently, Maladen \textit{et al}.\ \cite{maladen2009undulatory} studied the subsurface locomotion of sandfish in dry granular substrates. While the crawling and burying motion of a sandfish is driven by its limbs, an undulatory gait is employed for subsurface locomotion without use of limbs. Using high speed x-ray imaging, the subsurface undulating pattern of the sandfish body was found to be well described by a sinusoidal waveform. A major challenge in the quantitative analysis of locomotion in granular materials is a lack of validated force models like the Stokes equation in viscous fluids   \cite{zhang2014effective, goldman2014colloquium}. But inspired by the success of RFT for locomotion in viscous fluids, Maladen \textit{et al}.\ \cite{maladen2009undulatory} developed an empirical RFT in dry granular substrates for slender bodies (Sec.~\ref{subsec:RFT}), which was shown effective in modeling the undulatory subsurface locomotion of sandfish  \cite{maladen2009undulatory}. The proposed force model thus enables theoretical studies to address some fundamental questions on locomotion in granular media. In this paper we employ the proposed RFT to investigate the swimming characteristics of a slender filament of finite and infinite length undulating in a granular medium and compare the results with those in viscous fluids. In particular, previous analysis using the granular RFT considered only force balance in one direction  \cite{maladen2009undulatory} and hence a swimmer can only follow a straight swimming trajectory in this simplified scenario. Here we extend the results by considering a full three-dimensional force and torque balances, resulting in more complex kinematics such as pitching, drifting and reorientation. The swimming performance in relation to these complex kinematics is also discussed.

The paper is organized as follows. We formulate the problem and review the recently proposed RFT in granular media in Sec.~\ref{sec:form}. Swimmers of infinite length are first considered (Sec.~\ref{sec:inf}): we determine that the optimal waveform maximizing swimming efficiency, similar to results in viscous fluids, is a sawtooth (Sec.~\ref{subsec:opt}); we then study the swimming characteristics of sawtooth and sinusoidal swimmers in granular media and compare the results with swimming in viscous fluids (Sec.~\ref{subsec:sawNsine}). Next we consider swimmers of finite length (Sec.~\ref{sec:fin}) and characterize the effects of drifting and pitching in terms of propulsion speed and efficiency, before concluding the paper with remarks in Sec.~\ref{sec:discussion}.

\vspace{1.75in}

\section{Mathematical Formulation}
\label{sec:form}
\subsection{Kinematics}
\label{subsec:kinematic}
We consider an inextensible cylindrical filament of length $L$ and radius $r$ such that $r \ll L$, and assume that it passes a periodic waveform down along the body to propel itself in granular substrates. Following Spagnolie and Lauga \cite{spagnolie2010optimal}, the waveform is defined as $\bX(s) =[X(s),Y(s),0]^{\mathsf{T}}$, where $s \in [0,L]$ is the arc length from the tip. The periodicity of the waveform can then be described as
\begin{align}\label{eq:periodicity_condition}
X(s+\Lambda)=X(s)+\lambda, \quad Y(s+\Lambda)=Y(s),
\end{align}
where $\lambda$ is the wave length and $\Lambda$ the corresponding arc length along the body.  $N$ is the number of waves passed along the filament. Note that $L=N\Lambda$ and $\lambda =\alpha \Lambda$, where $0<\alpha<1$ is due to the bending of the body  \cite{spagnolie2010optimal}.

\begin{figure*}[!htb]
\centering
\includegraphics[scale=1]{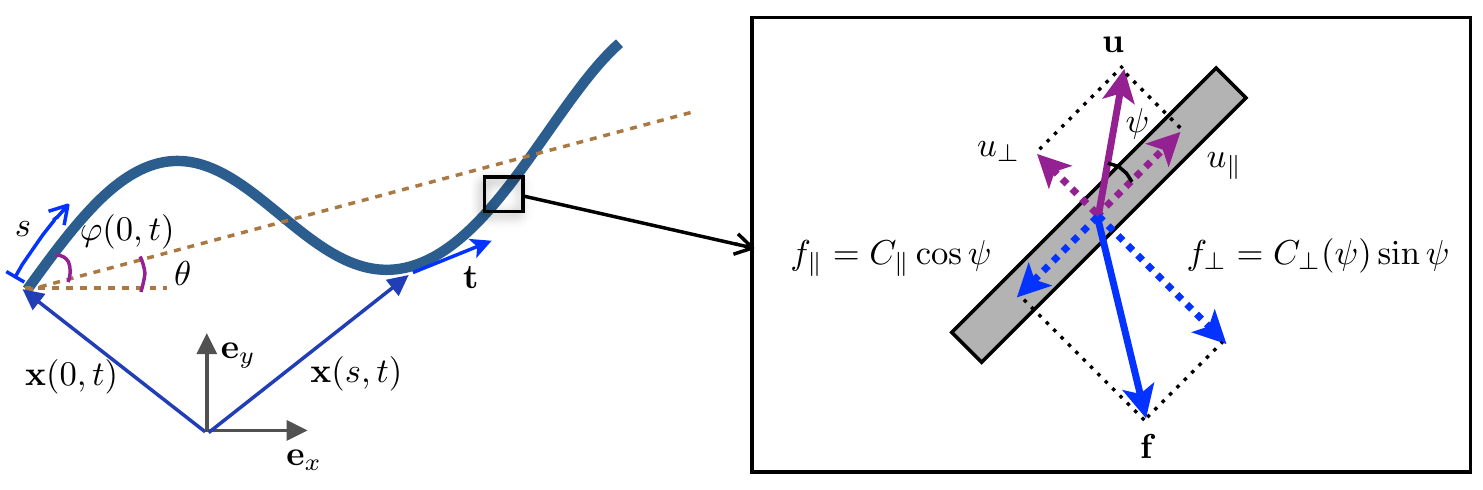}
\caption{\label{fig:schematic}Illustration of an undulating slender filament  and the resistive force theory in granular media. The body propagates a prescribed waveform to propel itself. Each element $\d s$ experiences a drag force $\d \bF = \bf \d s$. The basis vectors $\{\be_{x}, \be_{y}\}$ and the position vectors of its head  $\bx(0,t)$ and a material point $\bx(s,t)$ on the body in the lab frame are shown ($\be_{z} = \be_{x}\times \be_{y}$). The angle between the local velocity $\bu$ and unit tangent vector $\bt$ is $\psi(s,t)$.}
\end{figure*}

Initially, the filament is oriented along the $x$-axis of the lab frame with its head at $\bx_0$. At time $t$, the filament is passing the waveform at a phase velocity $\bV$ (with constant phase speed $V$) along the waveform's centerline, which is oriented at an angle $\theta(t)$ to the $x$-axis (Fig. \ref{fig:schematic}). In a reference frame moving with the wave phase velocity $\bV$, a material point on the filament is moving tangentially along the body with speed $c = V/\alpha$, and hence the period of the waveform is $T = \lambda/V = \Lambda/c$. By defining the position vector of a material point at location $s$ and time $t$  in the lab frame as $\bx(s,t)$,  we obtain
\begin{align}\label{eq:positionvec.}
\bx(s,t)-\bx(0,t)= \mathbf{\Theta}(t)\cdot\bR(s,t),
\end{align}
where
\begin{align}\label{eq:rotationmatrix}
\mathbf{\Theta}(t)=\begin{bmatrix}
	\cos\theta(t)&-\sin\theta(t)&0\\
	\sin\theta(t) &\cos\theta(t)&0\\
	0&0&1 \\	
	\end{bmatrix}
\end{align}
is the rotation matrix, and $\bR(s,t)=\bX(s,t)-\bX(0,t)$, and note that $\bX(s,t)= \bX(s-ct)$. Then, the velocity of each material point in the lab frame would be
\begin{align}\label{eq:velocity_relation}
\bu(s,t)=\dot{\bx}(0,t)+ \dot{\theta} \mathbf{\Theta}\cdot \bR^\perp+ \mathbf{\Theta} \cdot\dot{\bR} \textcolor{blue}{,}
\end{align}
where $\bR^\perp=\be_z \times \bR$, and dot denotes time derivative. The unit tangent vector in the direction of increasing $s$ is
\begin{align}\label{eq:tangent_vec}
\bt= \bx_s=\mathbf{\Theta}\cdot\bX_s(s,t),
\end{align}
where the subscript $s$ denotes the derivative with respect to $s$. The angle between the local velocity vector $\bu$ and the local unit tangent vector $\bt$ is $\psi$:
\begin{align}
\cos\psi = \buh\cdot\bt, \quad \buh = \frac{\bu}{\| \bu\|} \cdot
\end{align}

Now, to define the waveform we specify the tangent angle made with the centerline of the waveform
\begin{align}\label{eq:waveform}
\varphi(s,t)=\arctan\frac{Y_s}{X_s} \textcolor{blue}{\cdot}
\end{align}
Note that we have the following geometric relations:
\begin{align}
\bR&=\int_0^s \bX_s \d s, \quad \dot{\bR} = \int_0^s \dot{\varphi}\bX^\perp_s\d s,\label{eq:R}\\
\bt &= \mathbf{\Theta}\cdot\bR_s=\mathbf{\Theta}\cdot\bX_s,
\end{align}
where $\bX^\perp_s = \be_z\times \bX_s$,
and
\begin{align}
\alpha = \frac{\lambda }{\Lambda} = \frac{1}{\Lambda} \int_0^\Lambda \cos\varphi \d s.
\end{align}
The inextensibility assumption requires that $\partial [\bx_s\cdot\bx_s]/\partial t =0$, and the arc-length parameterization of the swimming filament naturally satisfies this constraint. The tangent angle is specified as a composition of different Fourier modes:
\begin{align}\label{eq:Fourier_psi}
\varphi(s,t)=\sum\limits_{n=1}^{n^*}  \left\{a_n \cos\left[\frac{2\pi n}{\Lambda}\left(s-ct\right)\right]+b_n \sin\left[\frac{2\pi n}{\Lambda}\left(s-ct\right)\right]\right\},
\end{align}
where
\begin{align}\label{eq:F_coef}
a_n&= \frac{2}{\Lambda}\int_0^\Lambda  \varphi(s,0) \cos\lb[\frac{2\pi n s}{\Lambda}\rb] \d s,\\
b_n&= \frac{2}{\Lambda}\int_0^\Lambda  \varphi(s,0) \sin\lb[\frac{2\pi n s}{\Lambda}\rb] \d s, \quad n=1, 2, 3, ...
\end{align}

\subsection{Resistive force theory}
\label{subsec:RFT}
In low Reynolds number swimming of a slender filament in a Newtonian fluids, the resistive forces are linearly dependent on the local velocity. The force per unit length exerted by the fluid on the swimmer body at location $s$ and time $t$ is given by
\begin{align}
\bf(s,t) = -K_T \bu \cdot\bt\bt -K_N(\bu-\bu\cdot\bt\bt),
\end{align}
where $K_N$ and $K_T$ are, respectively, the normal and tangential resistive coefficients. The self-propulsion of elongated filaments is possible because of drag anisotropy ($K_N \neq K_T $). A detailed discussion on this property can be found in the review paper by Lauga and Powers \cite{lauga2009hydrodynamics}. 
 Recent experimental studies of direct force and motion measurements on undulatory microswimmers in viscous fluids find excellent agreement with RFT predictions \cite{friedrich2010high,schulman2014dynamic}.
The ratio $r_{K}= K_N/K_T$ varies with the slenderness  ($L/r$) of the body. In the limit of an infinitely slender body, $L/r \rightarrow \infty$, $r_{K} \rightarrow 2$, which is the value adopted in this study.

For undulatory locomotion in dry granular media, we  only consider the slow motion regime where grain-grain and grain-swimmer frictional forces dominate material inertial forces \cite{maladen2009undulatory}. The motion of the swimmer is confined to the horizontal plane such that the change of resistance due to depth is irrelevant. In this regime the granular particles behave like a dense frictional fluid where the material is constantly stirred by the moving swimmer \cite{zhang2014effective}.  The frictional force acting tangentially everywhere on the surface of a small cylindrical element is characterized by $C_{F}$, which is refered to as the flow resistance coefficient \cite{maladen2009undulatory}. The other contribution to the resistive forces is the in-plane drag-induced normal force, which is characterized by $C_{S}$. Note that $C_{S}$ is a constant because the drag is independent of the velocity magnitude. The normal resistive coefficient $C_\perp$ depends on the orientation ($\psi$) of the element with respect to the direction of motion (Fig. \ref{fig:schematic}). In other words, the resistive force exerted by the granular material on the swimmer per unit length
\begin{align}\label{eq:RFTGM}
\bf(s,t)= -C_\parallel \buh\cdot\bt\bt-C_\perp(\buh-\buh\cdot\bt\bt),
\end{align}
where
\begin{gather}\label{eq:RFTco}
C_\parallel = 2rC_F,\\
C_\perp(\psi) = 2rC_F+ \frac{2rC_S \sin \beta_0}{\sin \psi} = C_\parallel \left(1+\frac{C_S\sin\beta_0}{C_F\sin\psi}\right),
\end{gather}
 $\tan \beta_{0}=\cot \gamma_{0} \sin \psi$ and  $\gamma_0$ is a constant related to  the internal slip angle of the granular media\textcolor{blue}{ \cite{maladen2009undulatory}}. 
Although a complete physical picture of the dependence of $C_\perp$ on the orientation $\psi$ remains elusive, the application of the granular RFT proves to be effective. Several studies have applied the granular RFT to study the locomotion of sand-swimming animals and artificial swimmers and found good agreement with experiments and numerical simulations  \cite{maladen2011undulatory,zhang2014effective}. A detailed discussion about the effectiveness of granular RFT on modelling sand-swimming can be found in a review article by Zhang and Goldman \cite{zhang2014effective}.

An important parameter characterizing the response of dry GM to intrusion is the volume fraction $\phi$, which is defined as the ratio of the total volume of the particles divided by the occupied volume. The level of compaction affects drag response as closely packed (high $\phi$) GM expands to flow while loosely packed (low $\phi$) material would consolidate \cite{maladen2009undulatory}. The drag parameters $C_{S}, C_{F}$ and $\gamma_{0}$ depend on the volume fraction of the GM. In our study, we refer to the GM with $\phi=0.58$ as loosely packed (LP) whereas $\phi=0.62$ as closely packed (CP). The numerical values of the drag parameters are adopted from the paper by Maladen \textit{et al}. \cite{maladen2009undulatory}, where the forces at a fixed depth of $7.62$ cm were measured by towing a cylinder of stainless steel.

Without external forcing, the self-propelled filament satisfies force-free and torque-free conditions:

\begin{align}
\bF& =\int_0^L \bf(s,t) \d s=\textbf{0}, \label{eq:Fbalance}\\
\bT &=\int_0^L [\bx(s,t)-\bx(0,t)] \times \bf(s,t)\d s= \textbf{0}.\label{eq:Tbalance}
\end{align}

The granular RFT exhibits the symmetry property that $\bu \to -\bu$ results in $\bf \to -\bf$. Combining this symmetry with the kinematics of the undulatory locomotion (see Sec.~\ref{subsec:kinematic}), one can show that the velocities $-\dot{\bx}(0,t)$ and $-\dot{\theta}$ are solutions to the instantaneous motion under a reversal of the actuation direction ($c\to -c$) provided that $\dot{\bx}(0,t)$ and $\dot{\theta}$ are solutions to the original problem (without reversal of the actuation). This symmetry is of course present in viscous RFT and this commonality, as we shall show, leads to qualitatively similar swimming behaviors.

\subsection{Swimming efficiency}
\label{subsec:efficiency}
The instantaneous swimming speed of the filament is given by $\dot{\bx}(0,t)$, and the mean swimming velocity is defined as $\bU = \left<\dot{\bx}(0,t)\right> = U_{x}\be_{x}+U_{y} \be_{y}$ with the magnitude $U=\lVert \bU\rVert$. The angle brackets $\left<...\right>$ denote time-averaging over one period $T$. The efficiency of the undulatory locomotion for a given deformation wave is defined by the ratio of the power required to drag the straightened filament through the surrounding substance to the power spent to propel the undulating body at the same velocity \cite{lighthill1975mathematica}. Hence, the efficiency for undulatory swimming of slender filaments in viscous fluid ($\eta_f$) and granular substance ($\eta_g$), respectively,  are
\begin{align}
\label{eq:effi}
\eta_f = \frac{K_T L U^2}{P}, \quad \eta_g = \frac{C_\parallel L U}{P},
\end{align}
where
\begin{align}
P = \left<\int_0^L \bf(s,t)\cdot\bu(s,t)\d s\right> \cdot
\end{align}
The optimal swimming can then be interpreted as either swimming with the maximum speed at a given power or swimming with the minimum power at a given speed.   
\subsection{Waveforms}
\label{subsec:waveform}
\begin{figure*}[!htb]
\centering
\includegraphics[scale=1]{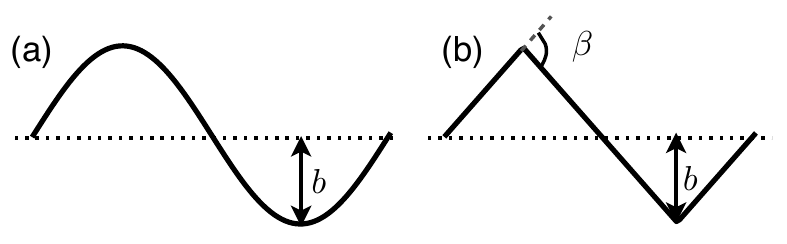}
\caption{\label{fig:wave} Undulating filaments with a single wave ($N=1$). (a): sinusoid, $kX_{0}=0$; (b) sawtooth, $kX_{0}=0$}
\end{figure*}

We consider two typical planar waveforms that have been well studied in Newtonian swimming: the sinusoidal waveform, and the sawtooth waveform (Fig.~\ref{fig:wave}). The sinusoidal waveform can be described by its Cartesian coordinates:
\begin{align}\label{eq:SineCart}
Y = b \sin k(X+X_{0}),
\end{align}
where $k = 2\pi/\lambda$ is the wave number, $kX_{0}$ is the initial phase angle of the waveform, and $b$ the wave amplitude. The dimensionless wave amplitude is defined as $\epsilon = kb$.

The sawtooth waveform, which consists of straight links with a bending angle $\beta$ ($\varphi = \pm \beta/2$), can be described as
\begin{align}
\label{eq:sawtoothEq}
Y = \frac{2b}{\pi} \arcsin[\sin k(X+X_0)],
\end{align}
The dimensionless amplitude $\epsilon = k b= (\pi/2)\tan(\beta/2)$.

\section{Bodies of infinite length}
\label{sec:inf}
For bodies of infinite length ($L\to\infty$), the swimming motion is steady and unidirectional, and hence $\dot{\theta(t})=0$. Without loss of generality, we assume the filament propagates the deformation wave in the positive $x$-direction. Then the velocity of a material point on the body can be written as
\begin{align}\label{eq:INFV}
\bu = -U \be_x +V\be_x-c\bt,
\end{align}
where $U$ is the swimming speed \cite{lighthill1975mathematica}. For an infinite swimmer, the unidirectional swimming velocity for a given waveform can be obtained from only the force balance in the $x$-direction, $\bF\cdot\be_x=0$, over a single wavelength,

\begin{align}
\int_0^\Lambda\left(\frac{C_S \sin\beta_0}{\sin\psi}+C_F\right)\buh\cdot\be_x\d s -\int_0^\Lambda\frac{C_S \sin\beta_0}{\sin\psi}(\buh\cdot\bt)\bt\cdot\be_x\d s=0.\label{eq:Fx0}
\end{align}
The above integral equation can be solved for $U$ numerically for a given waveform in general and but is analytically tractable in certain asymptotic regimes, which we discuss below.

\subsection{Optimal shape: numerical results}
\label{subsec:opt}
A natural question for swimming organisms is how their swimming gaits evolve under the pressure of natural selection  \cite{childress1981mechanics}, since being able to swim does not necessarily mean one does it efficiently. The understanding of optimal swimming may reveal nature's design principles  and guide the engineering of robots capable of efficient self-propulsion. As a response, the optimal strategies of several Newtonian swimming configurations have been studied. Becker \textit{et al}. \cite{becker2003self} determined the optimal strategy of Purcell's three-link swimmer under constant forcing and minimum mechanical work. Tam and Hosoi \cite{TamPRL} improved the swimming speed and efficiency of the optimal strategy of Purcell's three-link swimmer by allowing simultaneous rather than sequential movement of both hinges (kinematic optimization). Using viscous RFT, Lighthill showed that the optimal flagellar shape has constant angle between the local tangent to the flagellum and the swimming direction \cite{lighthill1976flagellar}. In 2D, the sawtooth profile with a tangent angle $\varphi \approx \pm 40^\circ$ (bending angle $\beta \approx 80^{\circ}$) was found to optimize the swimming efficiency of an infinite length swimming filament. Alternatively, this solution can be obtained through a variational approach \cite{spagnolie2010optimal}. In 3D, Lighthill's solution leads to an optimal shape of a rotating helix. More recently, Spagnolie and Lauga studied the optimal shapes for both finite and infinite elastic flagellum by incorporating physical constraints such as bending and sliding costs \cite{spagnolie2010optimal}. Inspired by the investigations of optimal strategies for Newtonian swimming, we study the optimal shape for infinite swimmers in granular substrates using resistive force theory. 

For bodies of infinite length, the optimal shape is time, scale  and phase invariant \cite{spagnolie2010optimal}. Therefore, we take $\Lambda=L=1$ and consider the optimization for $t=0$. In other words, the local tangent angle for the optimization problem would be
\begin{align}
\varphi(s,t=0)=\sum\limits_{n=1}^{n^*}  a_n \cos(2\pi ns).
\end{align}
We consider the optimal filament shape by maximizing the swimming efficiency $\eta$ defined in Sec.~\ref{subsec:efficiency}. Once the local tangent angle is obtained, the shape itself can be recovered by integration. The numerical methods used in this optimization can be found in the Appendix.

\begin{figure*}[!htb]
\centering
\includegraphics[scale=1]{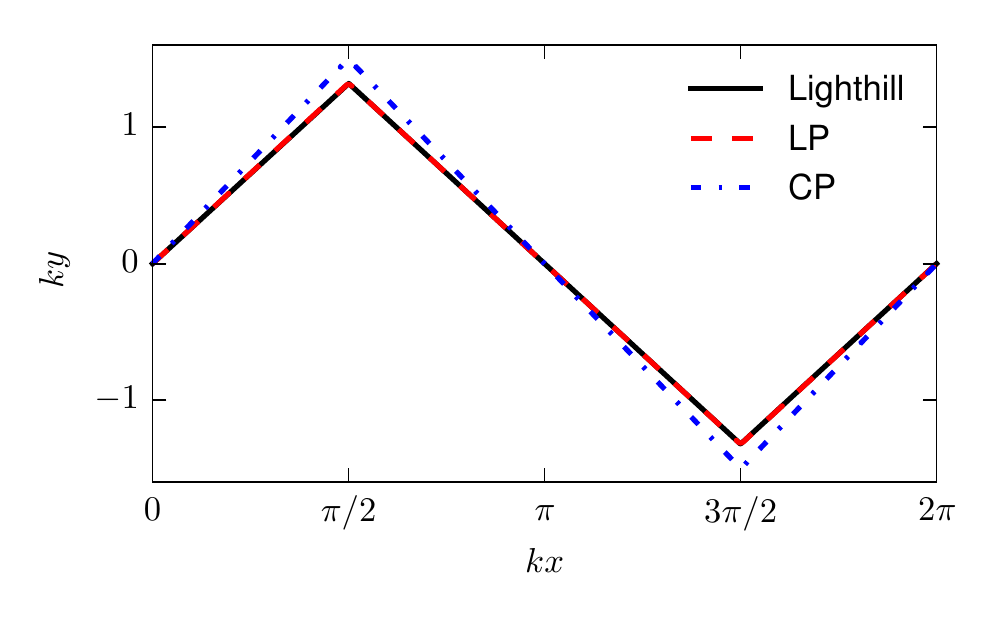}
\caption{\label{fig:opt}Optimal shapes in terms of swimming efficiency for an infinite filament in a granular substrate (LP, CP) and Newtonian fluid. The spatial coordinates are scaled to the same  wave length. For loosely packed granular material, the optimal shape is almost the same as the analytical result of Lighthill's in Newtonian fluid. }
\end{figure*}

The optimal shapes found by maximizing the swimming efficiency are presented in Fig.~\ref{fig:opt} for a LP granular substrate (red dashed line), a CP granular substrate (blue dash-dot line), and a viscous Newtonian fluid (black solid line) as a comparison. First, it is interesting that the optimal shape stays as sawtooth despite the nonlinearity in the resistive force model of granular substrates. 
The optimal bending angles for LP and CP granular media are, respectively, $\beta \approx 80^{\circ}$ and $\beta \approx 87^{\circ}$. The associated efficiencies of the optimal shapes are around $0.56$ for LP and $0.51$ for CP granular substrates, which are much greater than that of Newtonian swimming. In spite of the difference in the surrounding media, the optimal bending angle for granular substrates and viscous Newtonian fluids lie within the same range; in particular, the optimal sawtooth in LP closely resembles that in Newtonian fluids.

We argue that it is not surprising that the sawtooth waveform is optimal in both the viscous RFT and the nonlinear granular RFT. Given an angle that maximizes the efficiency of a local element. Without any penalty, the globally optimal shape would be the one that is locally optimal everywhere along the body. As a result, a local resistive force model should exhibit an optimal shape of a certain sawtooth waveform. Using this argument, we can simply drop the integration (or assume it is a sawtooth) in Eq. (\ref{eq:Fx0}) and consider the local optimality. The local optimal angle obtained is indeed the same as that found using numerical global optimization (see Sec.~\ref{subsec:sawNsine}).

The existence of a locally optimal tangent angle $\varphi$ originates from the physical picture introduced by the drag-based propulsion model \cite{lauga2009hydrodynamics} (Fig.~\ref{fig:schematic}). Let $\bu_{d}=u_{d}\be_{y}$ be the transverse deformation velocity of an infinite swimming filament. Then a propulsive force, which is perpendicular to the direction of the deformation velocity, generated by this deformation can be given by $\bf_{\textrm{prop}} = -(C_{\perp}(\psi)-C_{\parallel}) \sin\varphi\cos\varphi \be_{x}$. Therefore, the propulsive force arising from a local deformation of the filament scales with its orientation as $\sin\varphi\cos\varphi/\sqrt{\tan^{2}\gamma_{0}+\cos^{2}\varphi}$, the maximum of which is achieved when $\varphi \approx 64^{\circ}$. However, as the tangent angle increases, the power consumption of the swimming filament increases. As a result, the swimmer tends to reduce the tangent angle to decrease the energy expenditure while maintaining a relatively high propulsive force. It is the interplay of these two factors that determines the optimal tangent angle. 
 
\subsection{Sawtooth and sinusoid}
\label{subsec:sawNsine}
The swimming speed of an infinite sawtooth in viscous fluids can be expressed as
\begin{align}\label{eq:sawNewtoninf}
\frac{U}{V} = \frac{1-\cos\beta}{3-\cos\beta} \cdot
\end{align}

For a sawtooth profile in granular substrates, although an explicit analytical solution cannot be extracted, an implicit algebraic equation for the swimming speed $U$ can be obtained since the local resistive forces do not vary along the body:
\begin{align}\label{eq:SawtoothINF}
\left(\frac{C_S \sin\beta_0}{\sin\psi}+C_F\right)\buh\cdot\be_x -\frac{C_S \sin\beta_0}{\sin\psi}(\buh\cdot\bt)\bt\cdot\be_x=0,
\end{align}
where $\bt\cdot\be_x = \cos(\beta/2)$. We then solve Eq.~(\ref{eq:SawtoothINF}) numerically (see Appendix) with the same convergence criterion as in the optimization (Sec.~\ref{subsec:opt}). For a sinusoidal wave in granular media, a simplification like Eq.~\ref{eq:SawtoothINF} is not available and we therefore directly solve Eq.~(\ref{eq:Fx0}) with the numerical method outlined in the Appendix. 

\begin{figure*}[!htb]
\centering
\includegraphics[scale=1]{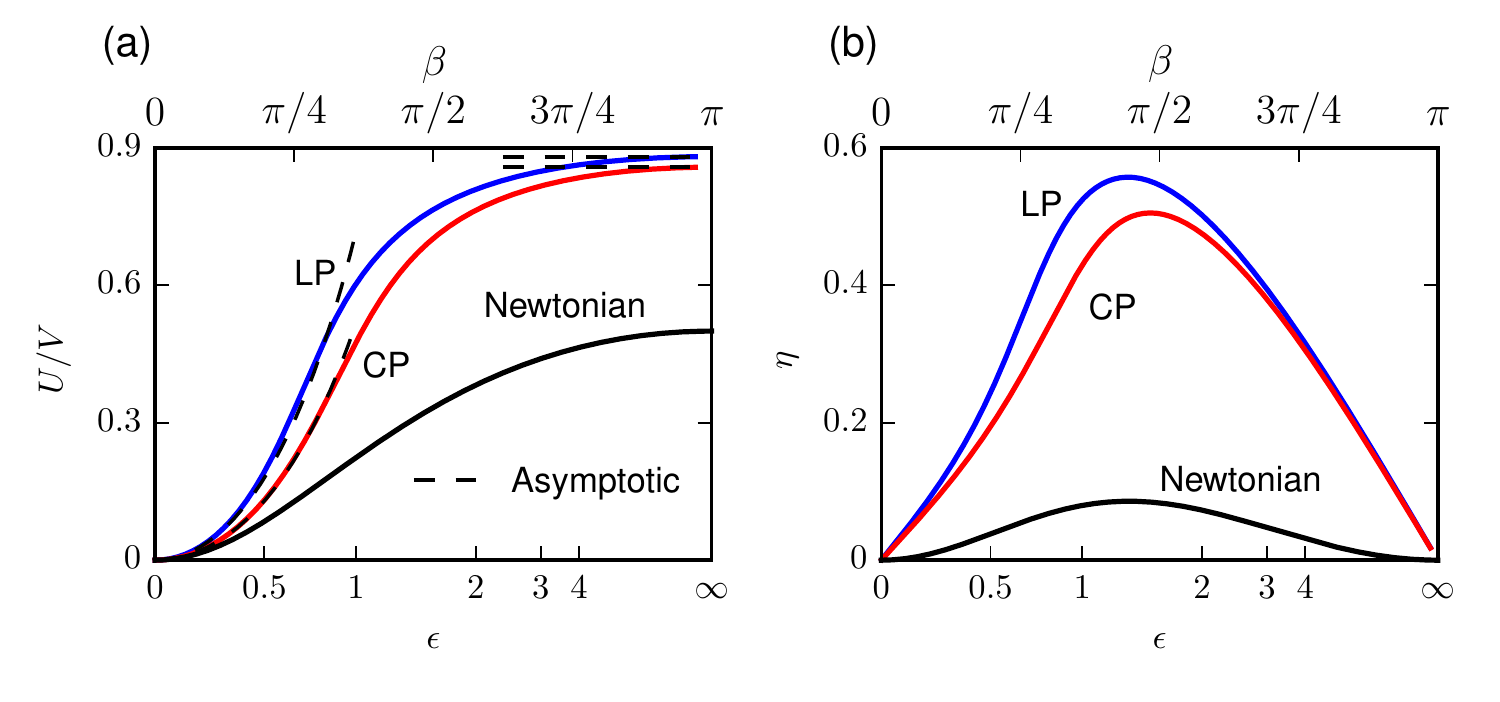}
\caption{\label{fig:inf}(a): Swimming speed of infinite sawtooth waveforms as a function of amplitude $\epsilon$ (or bending angle $\beta$) in granular material and Newtonian fluids. The dashed lines indicate the small and large amplitude asymptotic solutions. (b): Efficiency of infinite sawtooth waveforms as a function of amplitude $\epsilon$ (or bending angle $\beta$) in granular material and Newtonian fluids.}
\end{figure*}

For small amplitude sawtooth waveforms ($\epsilon \ll 1$), or small bending angle $\beta$, we obtain an asymptotic solution of the swimming speed $U$. Note that the swimming speed is invariant under a phase shift of $\pi$, which is equivalent to a sign change in the amplitude: $\epsilon \rightarrow -\epsilon$. Assuming a regular expansion in $\epsilon$, this symmetry argument leads to a quadratic scaling of the swimming speed in the wave amplitude \cite{pak2014theoretical}
\begin{align}
\frac{U}{V}\sim  \frac{4\cos\gamma_0C_S}{\pi^2C_F}\epsilon^2 \cdot
\end{align}
When the bending angle is large, another asymptotic limit can be obtained. The swimming speed $U/V$ approaches a constant as $\beta \to \pi$ and analytically we find that
\begin{align}
\frac{U}{V} \sim \frac{C_{S}}{C_{S}+C_{F}\tan\gamma_{0}} \cdot
\end{align}
One can also show that this large amplitude asymptotic limit for a sawtooth equals that of a sinusoidal wave. For small amplitude sinusoidal waveforms, however, the nonlinearity of the shape and the resistive forces results in a non-uniform integral and a slowly converging asymptotic series. To leading order, the swimming speed $U/V$ scales as  $\epsilon^{2}/\ln(1/\lvert\epsilon\rvert) $, which does not agree well with the numerical results even for $\epsilon< 0.1 $ as the higher order terms being truncated are not significantly smaller.  We present the small and large amplitude asymptotic solutions for the granular swimming of a sawtooth profile in Fig.~\ref{fig:inf}(a). The asymptotic solutions agree well with the numerical solutions even for wave amplitudes close to one. Fig.~\ref{fig:inf}(b) shows the efficiency of swimming as a function of the bending angle for an infinite sawtooth in both granular media and viscous fluids. For swimming efficiency, a global maximum in bending angle exists for both viscous and granular swimming. Note that the optimal angles obtained here are equal to those obtained via the global optimization (Sec.~\ref{subsec:opt}). 

\begin{figure*}[!htb]
\centering
\includegraphics[scale=1]{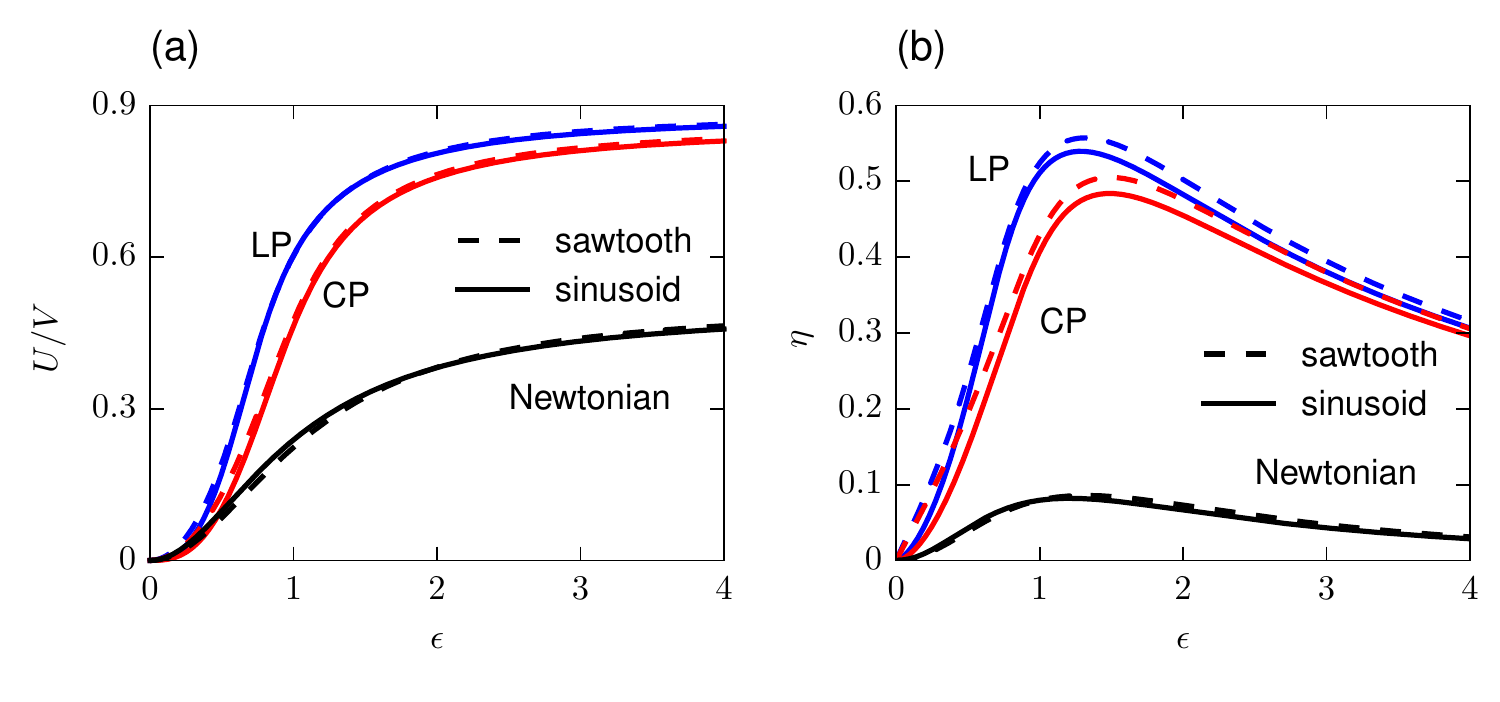}
\caption{\label{fig:sawNsine}A comparison of the swimming speed (a) and efficiency (b), as a function of wave amplitude $\epsilon$ for sawtooth and sinusoidal waveforms in granular substrates and Newtonian fluids.}
\end{figure*}

In Fig.~\ref{fig:sawNsine}, we compare the swimming speed and efficiency of sawtooth and sinusoidal waveforms in both GM and Newtonian fluids as a function of the wave amplitude $\epsilon$. In both GM and Newtonian fluids, the swimming speed of a sawtooth  is only slightly different from that of a  sinusoid with the same dimensionless amplitude. This small difference indicates that the effects of the local curvature variations are not significant in both the granular and viscous RFT. Although the sawtooth is found to be the mathematically optimal shape, the undulatory gait of a sandfish resembles a smooth sinusoidal waveform  \cite{maladen2009undulatory}. The slight difference in swimming performance between the two waveforms presented in this section might justify the adoption of a sinusoidal waveform instead of the mathematically optimal sawtooth waveform, since the kinks in the sawtooth may involve other energetic costs associated with bending and the deformation of the internal structure of the body  \cite{spagnolie2010optimal}.

\section{Bodies of finite length}
\label{sec:fin}
The infinite swimmer model only enforces a force balance in one direction and hence a swimmer is confined to swim only unidirectionally without any rotation. In reality, however, a swimmer has a finite size and more complex swimming kinematics, including transverse motion relative to the wave propagation direction and rotation. Previous studies employed slender body theory to investigate the swimming motion of finite filaments in a viscous Newtonian fluid and their swimming performance in relation to number of wavelengths and filament length  \cite{pironneau1974optimal, higdon1979hydrodynamic, spagnolie2010optimal,koehler2012pitching}. In this section, we investigate the swimming characteristics of finite-length sinusoidal swimmers in a granular medium and compare with their Newtonian counterparts. The numerical methods implemented to solve the equations of motion of a finite length swimmer are given in the Appendix.

\subsection{Geometries}
\label{subsec:geom}

\begin{figure*}[!htb]
\centering
\includegraphics[scale=1]{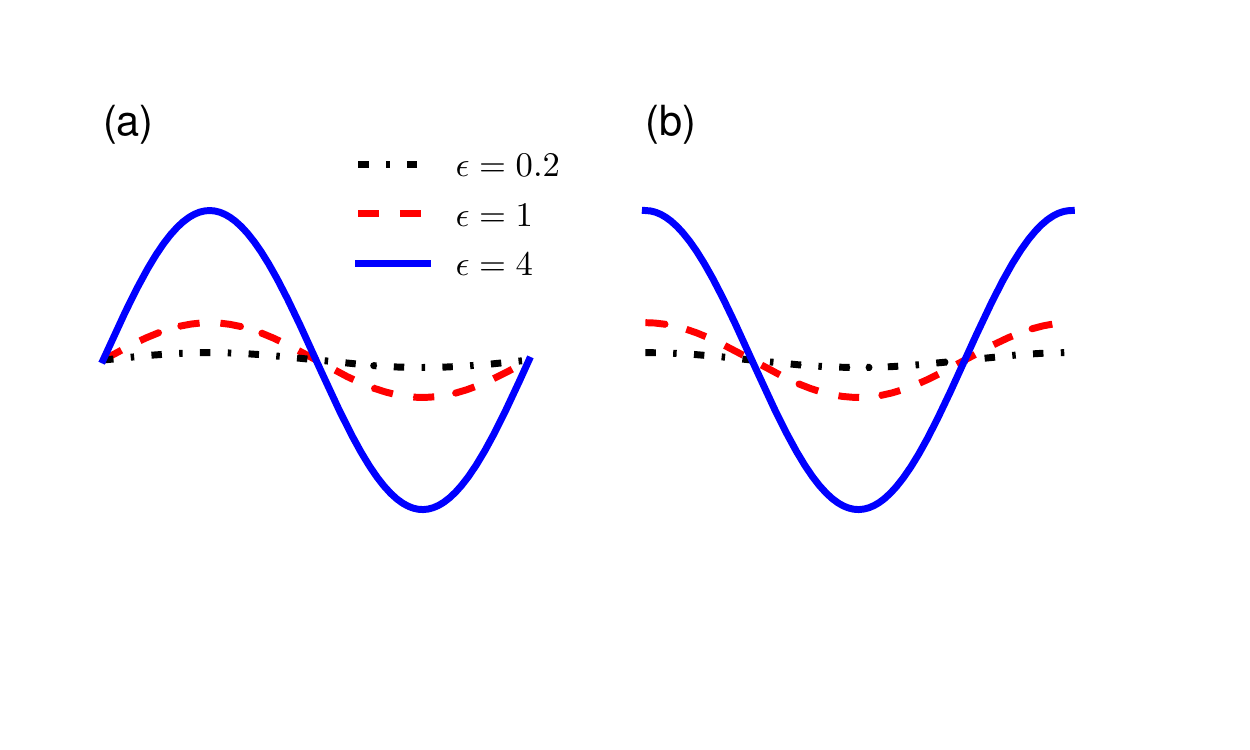}
\caption{\label{fig:oddEven}Shapes of swimming finite length single wave ($N=1$) sinusoidal filaments for different wave amplitude $\epsilon$. (a): the odd sine configuration, with $kX_{0}=0$, (b): the even cosine configuration, with $kX_{0}=\pi/2$. The waveforms are rescaled to the same wave length for better comparison.}
\end{figure*}

For an undulating sinusoidal filament, the initial shape of the swimmer is determined by the number of waves $N$, the wave amplitude $\epsilon$, and the initial phase angle $kX_{0}$ (Eq.~(\ref{eq:SineCart})). The two specific categories of shapes that possess odd or even symmetry for a single wave sinusoidal swimmer are shown in Fig.~\ref{fig:oddEven}. A swimmer in an odd configuration is the one that has point symmetry about the midpoint of the filament as seen in Fig.~\ref{fig:oddEven}(a), while an even configuration is the one that possesses mirror symmetry about the vertical line through the midpoint as in Fig.~\ref{fig:oddEven}(b). In our paper, the shapes shown in Fig. \ref{fig:oddEven}(a) are referred to as odd sine swimmers, while even cosine swimmers are those shown in Fig. \ref{fig:oddEven}(b). Note that an even sine swimmer would be the one that has the number of waves $N \in \{1/2, 3/2,5/2, ...\}$ and a phase angle $kX_{0} \in \{ 0,\pm \pi, \pm2\pi, ...\}$; an even cosine swimmer is the one that has the number of waves $N \in \{1, 2, 3, ... \}$ and a phase angle $kX_{0}\in \{\pm\pi/2, \pm3\pi/2, ... \}$.

\subsection{Pitching, drifting and reorientation}
Unlike the swimming of an infinite length undulatory swimmer whose motion is steady and unidirectional, the locomotion of a finite filament may also experience net motion normal to the initial direction wave propagation direction, also referred to as drifting, and unsteady rotational motion, known as pitching. Here we characterize in GM the re-orientation of a finite swimmer that results in drifting, and the dependence of swimming performance on pitching motion, previously reported to diminish performance in viscous Newtonian media \cite{spagnolie2010optimal,koehler2012pitching}.

\begin{figure*}[!htb]
\centering
\includegraphics[scale=1]{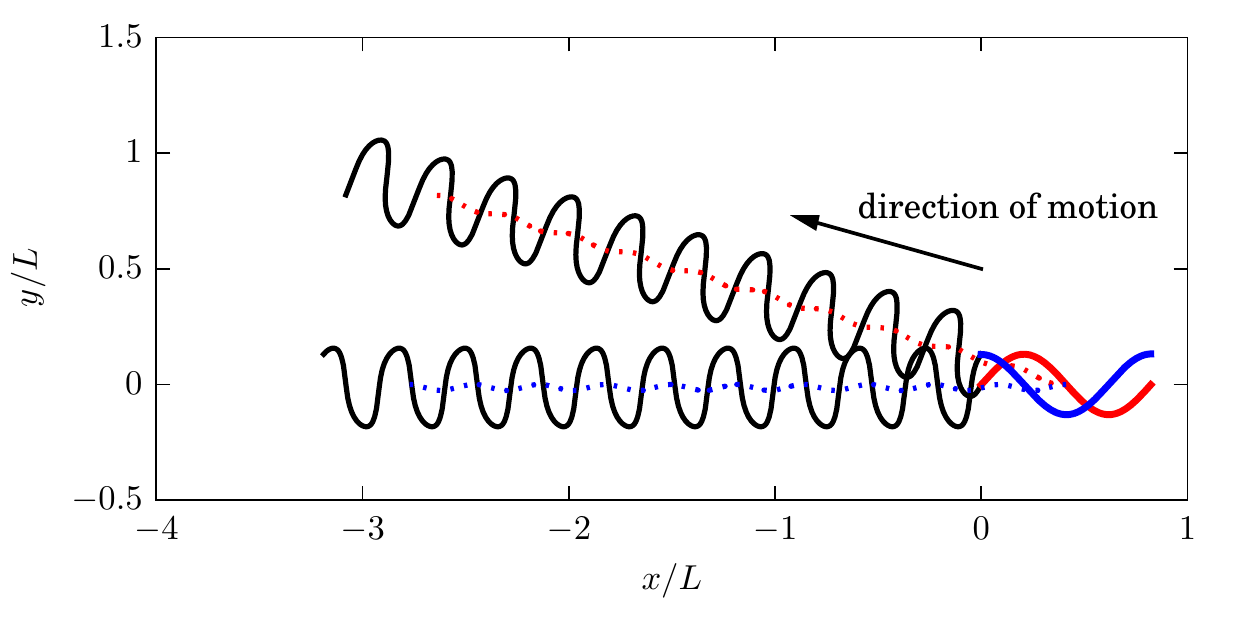}
\caption{\label{fig:traj}Trajectory of the head $\bx(0,t)$ (black solid lines) and trajectory of the swimmer centroid (dotted lines)  for swimming finite sinusoidal filaments with $N=1$ and $\epsilon =1$ that possess odd/even symmetry at $t=0$ in loosely packed GM. The filament swims towards the left when the wave propagates to the right. If the configuration possesses even symmetry it does not undergo a net reorientation.}
\end{figure*}

For an even symmetry filament in viscous fluids, Koehler \textit{et al}. \cite{koehler2012pitching} showed that the velocity of the center of mass is along the centerline of the waveform, hence the net drifting is zero. This argument relies on the kinematic reversibility of Stokes flow: reflection about the vertical line is equivalent to a time reversal (or reversing the direction of the actuation), so the instantaneous swimming is identical to the mirror reflection of its time-reversal, and the linearity requires the reverse of velocity due to time-reversal, thus one can show that the transverse component of the velocity is zero. As a result, the net displacement in one period for a filament starts with the even configuration is along the initial waveform centerline.

Although the granular RFT is nonlinear, the aforementioned symmetry property ($\bu \to -\bu \Rightarrow \bf \to -\bf$, see Sec.~\ref{subsec:RFT}) means that the same argument for an even symmetry swimmer can be made in GM. Therefore, zero net transverse motion is achieved if the swimmer starts with an even symmetry, which is also corroborated by the numerical simulation. Fig.~\ref{fig:traj} shows the head trajectories of two swimming sinusoidal filaments with the same wave amplitude ($\epsilon=1$), one starts with even symmetry while the other starts with odd symmetry. The net displacement of the even cosine swimmer is in the negative $x$-direction, which is the opposite direction of the wave propagation at $t=0$.  The odd sine swimmer, however, appears to be drifting upwards to the positive $y$-direction through time. 

\begin{figure*}[!htb]
\centering
\includegraphics[scale=1]{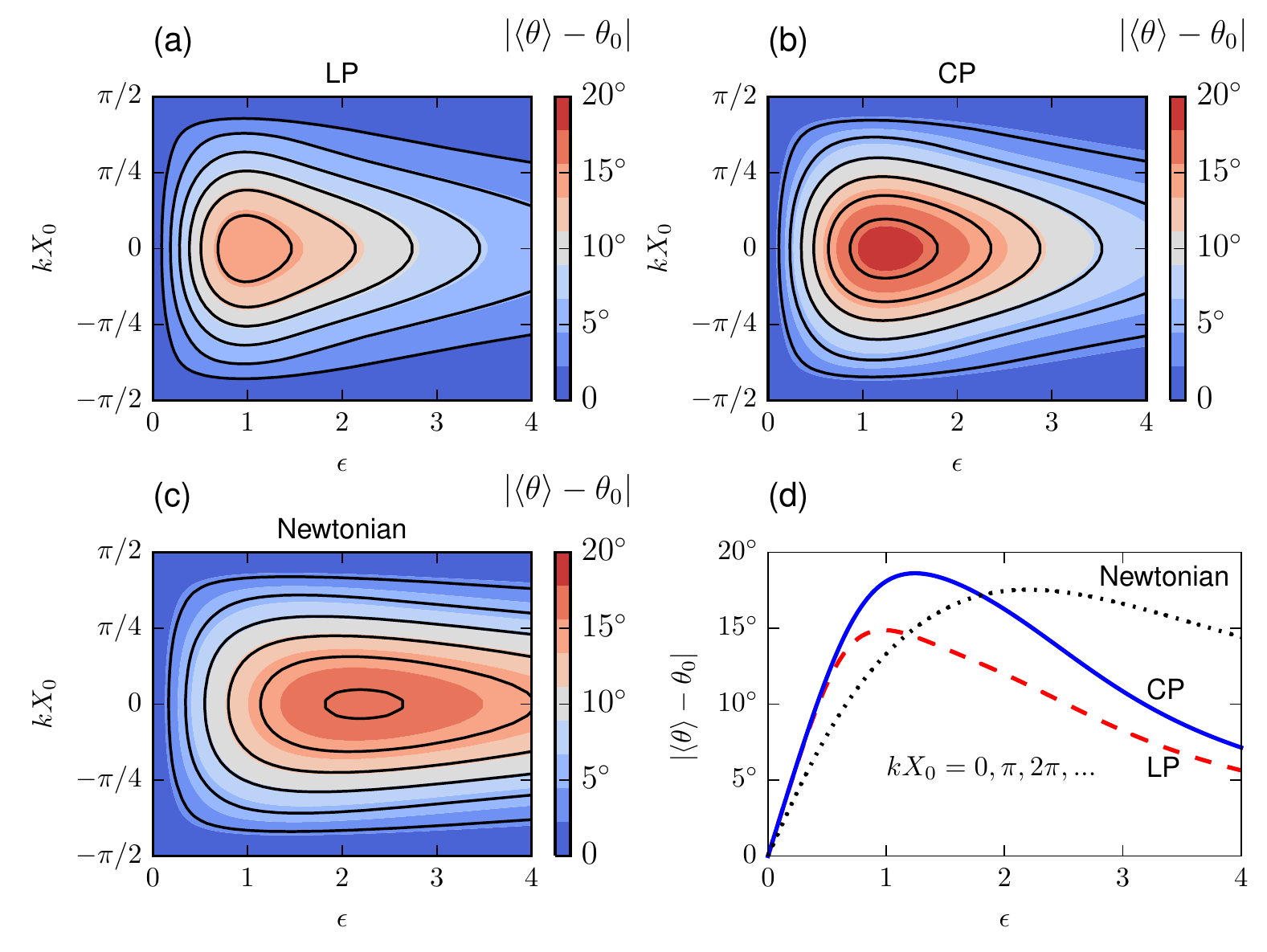}
\caption{\label{fig:theta}Parametric plots for the magnitude of the reorientation angle $\lvert\left<  \theta \right>-\theta_0\rvert$ for a single wave ($N=1$) sinusoid in (a) loosely packed GM, (b) closely packed GM and (c) Newtonian fluids. (d): Plots of $\lvert\left<  \theta \right>-\theta_0\rvert$ against the wave amplitude $\epsilon$ for the odd sine configuration in GM and Newtonian fluids. $\lvert\left<  \theta \right>-\theta_0\rvert$ is periodic with a period  of $\pi$.}
\end{figure*}

The swimming behavior presented in Fig. \ref{fig:traj} can be understood by examining the periodic instantaneous motion of the swimmer. In the moving frame, or the Lagrangian frame, the instantaneous motion of the swimmer can be viewed as being pulled through a waveform-shaped tube \cite{koehler2012pitching}. This motion, in turn, causes rotation and translation of the Lagrangian frame. The instantaneous rotation of the Lagrangian frame is described by $\theta(t)$, which is periodic due to the periodicity of the wave propagation. The average of $\theta(t)$ over one period, denoted as $\left< \theta\right>$, describes the average swimming direction. This angle $\left< \theta \right>$ is the same in every period which results in a straight line trajectory on average. If a filament, starts with an odd (even) configuration at $t =0$ (if  aligned with the $x$-axis then $\theta_0=0$), it would possess even (odd) symmetry at $t = T/4$. Thus the filament alternates between even symmetry and odd symmetry after successive time steps of $T/4$. In this viewpoint, $\left<\theta\right>-\theta_0$ characterizes the amount of time $t_{1}$ required for the filament to reorient itself such that it reaches an even symmetry. After that, the swimmer would move in the direction of the waveform centerline at $t=t_{1}$. For a fixed number of waves $N$ and amplitude $\epsilon$, the odd configuration requires the largest amount of time ($T/4$) to reach an even symmetry, therefore has the largest angle of reorientation. Note that the angle of reorientation should be distinguished from pitching of the swimmer, which is the instantaneous rotation of the swimmer about its waveform centerline.

In Fig.~\ref{fig:theta}, we present parametric plots of absolute value of the angle of reorientation $\lvert\left< \theta \right>-\theta_0\rvert$ by varying the wave phase angle $kX_{0}$ and the amplitude $\epsilon$ in both GM and viscous fluids. The number of waves is fixed as $N=1$, which approximates the shape of an undulating sandfish body \cite{maladen2009undulatory}. Note that a phase shift of $\pi$ would result in  a reversal of the direction of the transverse motion, hence the sign of  $\left<\theta\right>-\theta_0$. In both GM and Newtonian fluids, the maximum in $\lvert\left< \theta \right>-\theta_0\rvert$ is obtained when the filament possesses an odd symmetry at $t=0$, i.e., $kX_{0} \in\{ 0, \pi, 2\pi, ...\}$. For shapes that possess even symmetry, namely,  $kX_{0} \in\{ \pi/2, 3\pi/2, ...\}$, zero  transverse motion is observed. Within our parameter range, a maximum in $\lvert\left< \theta \right>-\theta_0\rvert$ is achieved around an intermediate value of the amplitude for a given phase angle. As an example, the variation of  $\lvert\left< \theta \right>-\theta_0\rvert$ with the amplitude $\epsilon$ for the odd configuration is shown in Fig.~\ref{fig:theta}(d). The largest amount of reorientation of an odd swimmer is achieved when $\epsilon \approx 1-1.2$ in GM while $\epsilon \approx 2.2$ in viscous fluids. We also note that the angle of reorientation decreases with the increasing of wave amplitude in the large amplitude region ($\epsilon>2$).

\begin{figure}[!tb]
\centering
\includegraphics[scale=1]{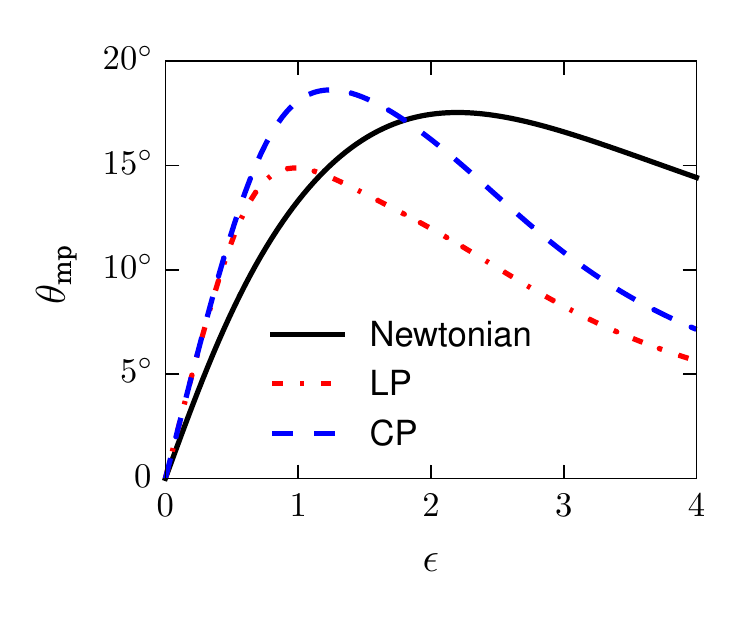}
\caption{\label{fig:tm} Maximum instantaneous pitching angle $\theta_{\textrm{mp}}$  as a function of the wave amplitude $\epsilon$ for single wave ($N=1$) sinusoidal swimmers  in GM and Newtonian fluids. }
\end{figure}

Although the transverse motion of the even configuration is minimal, the instantaneous pitching, $\theta(t)-\lb<\theta\rb>$, which generally diminishes performance,  can be significant. Multiple metrics  have been used to characterize  pitching of a swimmer \cite{koehler2012pitching,spagnolie2010optimal}, here we use the maximal amount of instantaneous pitching a swimmer can experience in one cycle of its motion $\theta_{\textrm{mp}} = \lvert\theta(t)-\left< \theta \right>\rvert_{\max}$. Fig.~\ref{fig:tm} shows the maximal instantaneous pitching angle $\theta_{\textrm{mp}}$ for single wave sinusoidal swimmers  in GM and Newtonian fluids. The maximal instantaneous pitching angle of a single wave sinusoid goes up to about $15^{\circ}$ in loosely packed GM while around $19^{\circ}$ in closely packed GM.

The instantaneous pitching of the swimmer results in a tortuous motion with a net swimming speed smaller than that of an infinite sinusoid. For a fixed number of waves and wave amplitude, a phase shift only leads to a variation in the direction of swimming. In other words, the velocity magnitude $U$ is independent of $kX_{0}$ but the $x$ and $y$ components vary. From a control point of view, one can change the phase angle of an artificial sinusoidal swimmer to obtain the desired direction of swimming.

\subsection{Swimming performance}

The two typical metrics for swimming performance used in the literature are the dimensionless swimming speed  $U/V$ and the swimming efficiency $\eta$, see Eq. (\ref{eq:effi}). For a sinusoidal swimmer, the performance depends on the dimensionless amplitude $\epsilon$ and the number of waves $N$. Note that the initial phase angle $kX_{0}$ does not affect the two performance metrics. The desired motion of a finite swimmer is its translation, therefore the optimization of a finite sinusoidal filament requires minimizing pitching.

\begin{figure}[!tb]
\centering
\includegraphics[scale=1]{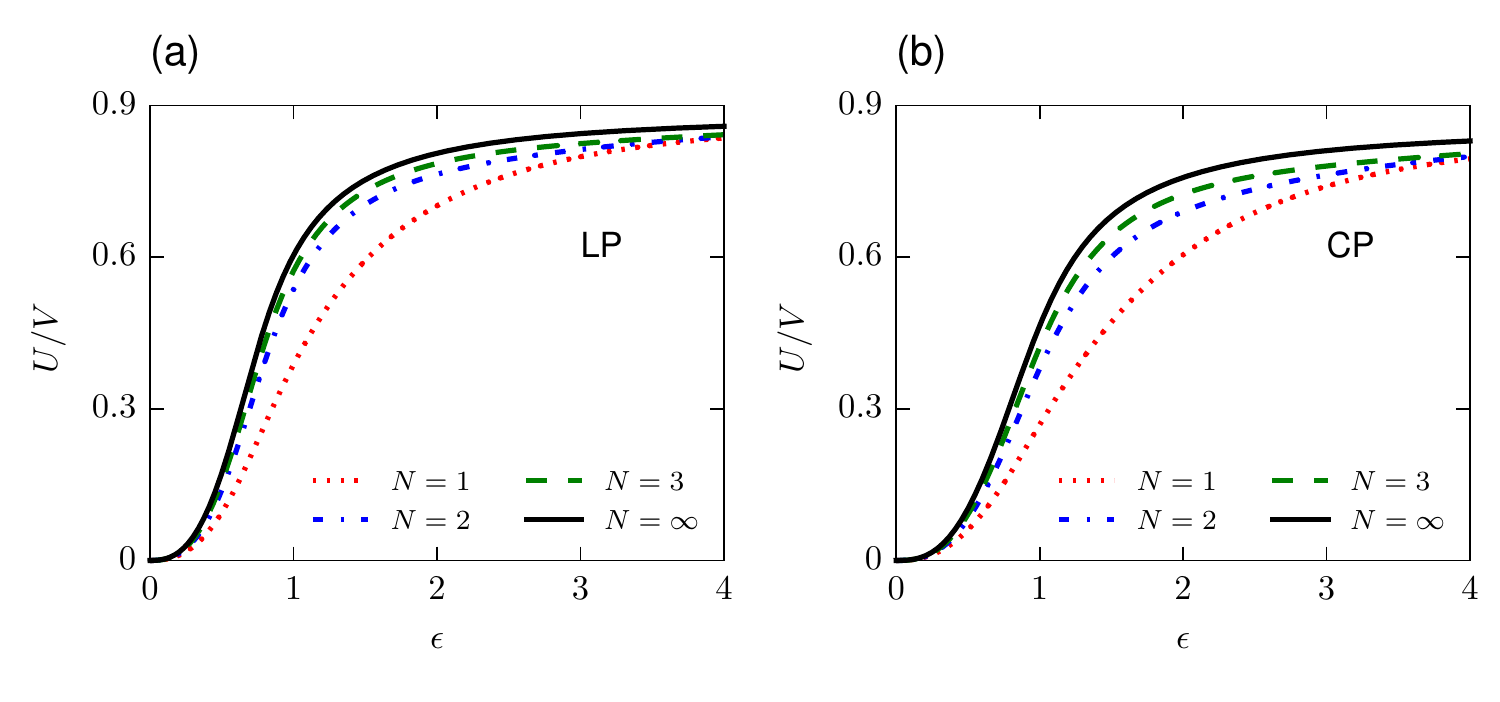}
\caption{\label{fig:UNGV}Swimming speed $U/V$ as a function of the dimensionless amplitude $\epsilon$ for different number of waves $N$ in (a) loosely packed GM and (b) closely packed GM.  The solid lines denote the swimming speed of an infinite sinusoid.}
\end{figure}

\begin{figure}[!tb]
\centering
\includegraphics[scale=1]{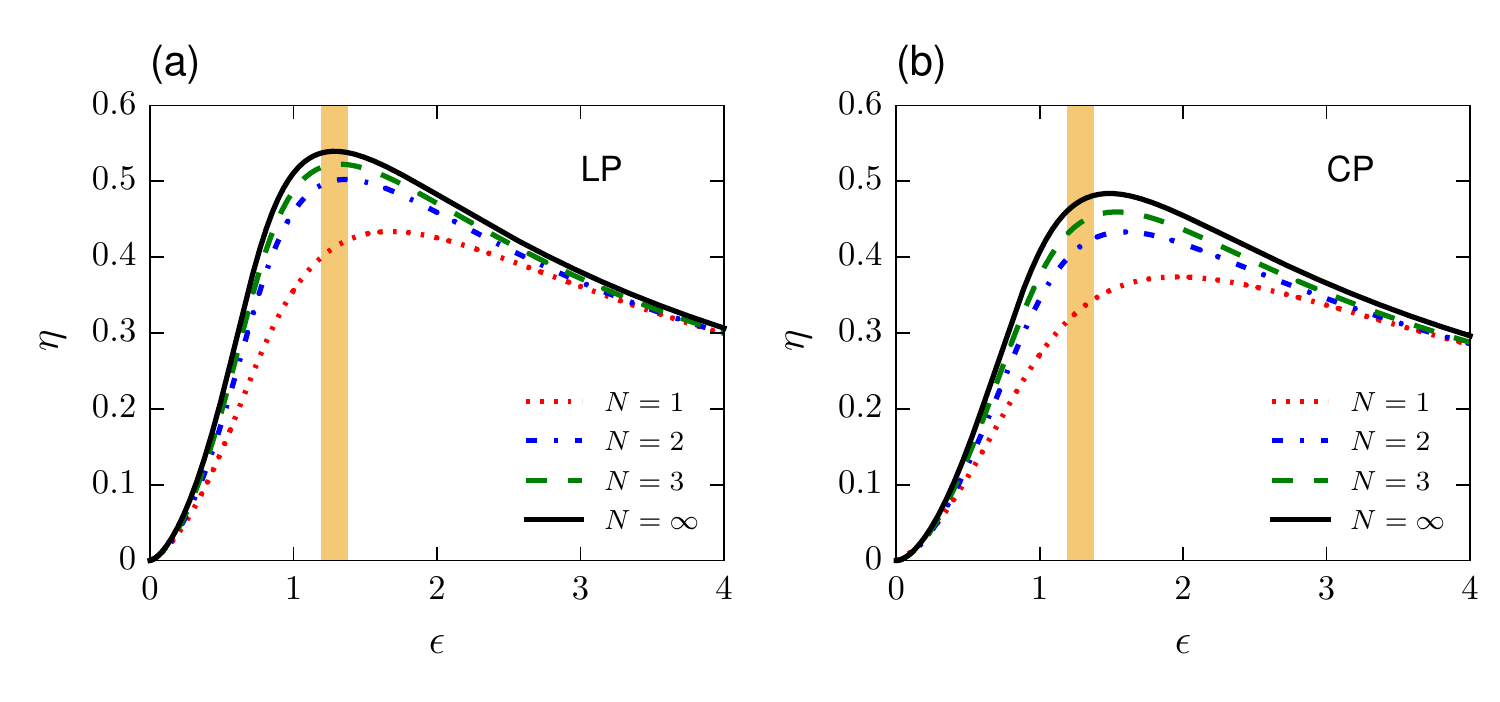}
\caption{\label{fig:efi}Swimming efficiency $\eta$ as a function of the dimensionless amplitude $\epsilon$ for different  number of waves $N$ in (a) loosely packed GM and (b) closely packed GM. The shaded regions represent the observed values of $\epsilon$ for lizards reported in the literature  \cite{maladen2009undulatory, ding2012mechanics}.}
\end{figure}

For an undulatory finite filament in viscous fluids, several studies have characterized the swimming performance and optimal strategies. Spagnolie and Lauga reported that the local maxima in swimming efficiency occur for around half-integer number of waves ($N \approx 3/2, 5/2, ...,$) when the bending cost is small \cite{spagnolie2010optimal}. Later studies by Koehler \textit{et al}. \cite{koehler2012pitching} and Berman \textit{et al}. \cite{berman2013undulatory} also showed that, for a sinusoidal swimmer,  local maxima in performance are achieved for close to half-integer number of waves where pitching is small.

We first verify that the swimming velocity (Fig.~\ref{fig:UNGV}) and efficiency (Fig.~\ref{fig:efi}) of a finite sinusoidal swimmer in GM both converge to that of an infinite sinusoidal swimmer as the number of waves $N$ increases. For a single wave sinusoid ($N=1$) in loosely packed GM, the optimal dimensionless amplitude that maximizes the efficiency is $\epsilon\approx 1.68$. As the number of waves increases, the optimal dimensionless amplitude approaches that of an infinite sinusoid ($\epsilon \approx 1.33$). Similar observations can be made for closely packed GM. We also observe that for a given dimensionless amplitude $\epsilon$, the difference in the swimming velocity (or efficiency) between a short swimmer ($N=1$) and an infinite swimmer can be associated with the pitching motion: the largest difference in swimming speed (or efficiency) between the $N=1$ and $N=\infty$ swimmers occurs in the region $\epsilon \approx 1$ in Figs.~\ref{fig:UNGV} and \ref{fig:efi}, which is also the region where pitching is the most significant (Fig.~\ref{fig:tm}).

\begin{figure}[!tb]
\centering
\includegraphics[scale=1]{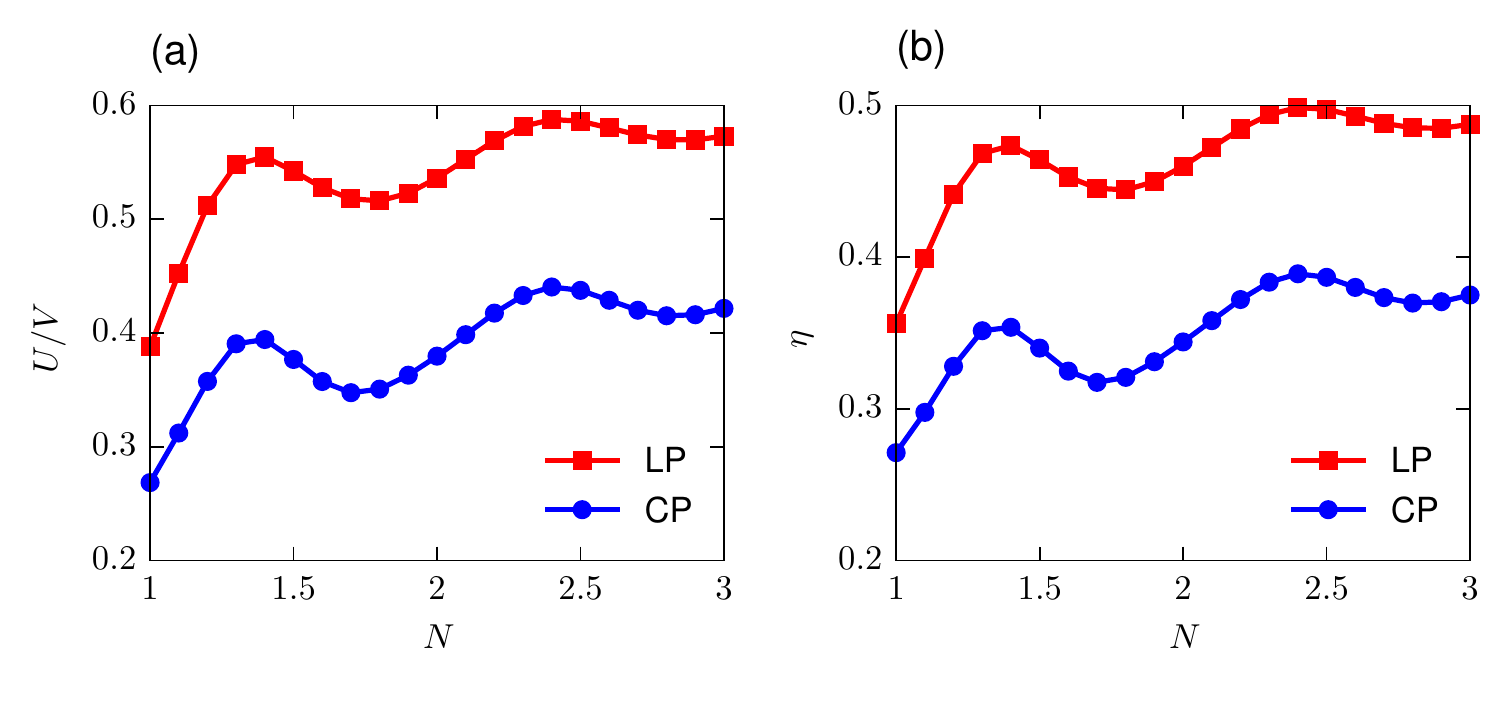}
\caption{\label{fig:N} (a) Swimming speed as a function of the number of waves in GM. (b) Swimming efficiency as a function of the number of waves in GM. The dimensionless amplitude is fixed ($\epsilon=1$).}
\end{figure}

\begin{figure}[!tb]
\centering
\includegraphics[scale=1]{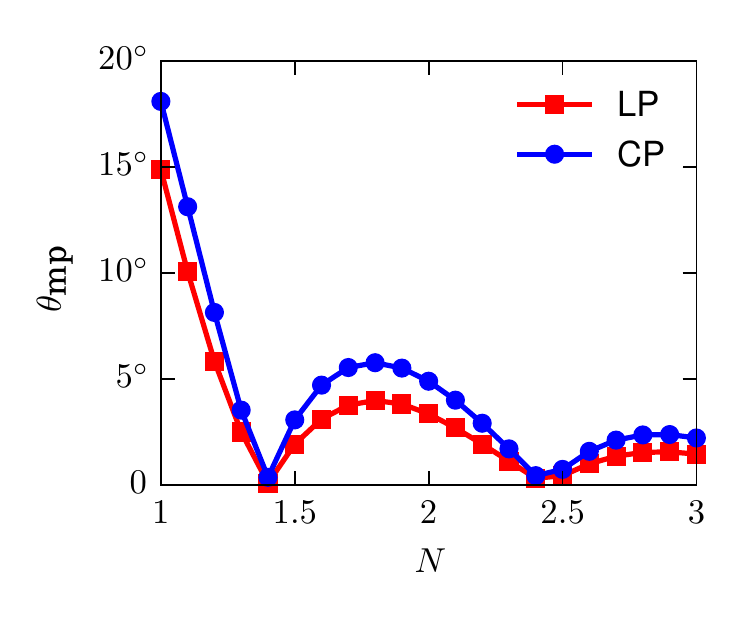}
\caption{\label{fig:tmp} Maximum instantaneous pitching angle as a function of the number of waves in GM. The dimensionless amplitude is fixed ($\epsilon=1$).}
\end{figure}

For a given waveform, the amount of pitching can be altered by changing the number of waves $N$. We investigate in Fig.~\ref{fig:N} the dependence of the performance metrics on the number of waves for a finite sinusoidal swimmer, keeping dimensionless amplitude fixed at $\epsilon=1$. Rather than approaching the swimming velocity (or efficiency) of the corresponding infinite sinusoid monotonically with increasing number of waves, the swimming speed and efficiency exhibit local maxima and minima. Similar to the Newtonian case, the local maxima in efficiency and swimming speed occur for the number of waves close to (but not equal) half-integers. The volume fraction of the GM has no significant influence on the number of waves where local maxima in swimming performance occur. As shown in Fig. \ref{fig:N}, the first local maximum in swimming performance for the number of waves greater than one occurs around $N\approx1.4$. The maxima in swimming performance are associated with minimal pitching as shown in Fig.~\ref{fig:tmp}. Finally we note that although both the first two local maxima have minimal pitching (Fig.~\ref{fig:tmp}), the swimmer with more number of waves ($N \approx 2.5$) still displays better swimming performance, which can be attributed to a smaller bobbing motion \cite{koehler2012pitching} (the relative motion of the center of mass of the swimmer to the net swimming direction) for the swimmer with more number of waves.

Finally, we relate our findings to biological observation; we show, in the shaded regions of Fig.~\ref{fig:efi}, the observed dimensionless amplitude (amplitude-to-wavelength ratio) for lizards reported in the literature ($\epsilon = 1.20-1.38$)  \cite{maladen2009undulatory, ding2012mechanics}. We see in the case of both loosely-packed and closely-packed granular media, that the biologically observed range of wave amplitudes sample high efficiencies not far from optimal ($\epsilon \approx 1.69$ for LP and $\epsilon\approx 1.95$ for CP for $N=1$). Since the efficiency peak is broad, a swimmer may adopt a close-to-optimal shape at the expense of only a modest drop in swimming efficiency to address other constraints (such as bending costs or internal dissipation).

\section{Conclusion} \label{sec:discussion}
In this paper, we have investigated locomotion of slender filaments in granular media using a resistive force theory proposed by Maladen \textit{et al}. \cite{maladen2009undulatory}. While previous work focused on infinite swimmers (or 1-D swimming) in reality a swimmer has a finite size, which leads to more complex swimming motion. By taking into account full force and torque balances, a finite swimmer is no longer only confined to swim in a straight trajectory. The orientation of the swimmer can be controlled by adjusting the features of the waveform such as the amplitude, phase, and number of wavelengths, allowing a swimmer to move from an initial position to a final destination via a more complex, designated trajectory. These degrees of freedom enable the control of swimmers without the use of any external fields to actively steer the swimmer. Our studies characterize this complex swimming motion in granular media, which may be useful for the development of programmable and efficient autonomous locomotive systems in such environments, but also suggest that swimmers in nature are themselves closely tuned for optimality.

We also find that undulatory locomotion of filaments in granular media is distinctly similar to that in viscous fluids. We compared a number of observations made for swimming in viscous fluids with RFT both for finite and infinite swimmers and found qualitatively similar behavior using granular resistive force theory despite the nonlinearity of the force law. The reason is largely down to two distinct similarities. The first, is that both laws are still local and thus ignore interactions of distinct parts of the body through the medium in which they swim. Ultimately this leads to finding that a sawtooth profile optimizes locomotion in both viscous fluids and granular media. The second, is that both force laws display the symmetry that $\bu\to-\bu$ results in $\bf\to-\bf$. This leads to a kinematic reversibility in both cases, where a reversal of the wave speed leads to an reversal of the translational and rotational motion of the swimmer, and hence a myriad of qualitatively similar behaviors that we have explored and quantified in the paper. 

\begin{acknowledgments}
Funding (to GJE) from the Natural Science and Engineering Research Council of Canada (NSERC) is gratefully acknowledged.
\end{acknowledgments}

\newpage

\appendix*
\section{Numerical implementation} % (fold)
\label{sec:appendix.}
In this appendix, we present the numerical methods implemented in the optimization of infinite filaments and the solution to the equations of motion of finite length filaments. 

\subsection{Optimization}
\label{subsec:APPENDIX_opt}
The numerical optimization (see Sec.~\ref{subsec:opt}) is performed using MATLAB's built-in \textit{fminsearch} function, which implements the Nelder-Mead simplex algorithm. We truncate the Fourier series by taking $n^*=100$ to have a sufficient spectral accuracy and use $m=1000$ points for the Gauss-Legendre integration scheme. Further increase in spectral and spatial resolution has a negligible effect on the optimization. The optimization search routine iterates until the algorithm detects a local solution gradient with a relative error tolerance of $10^{-14}$. For each iteration, the swimming speed $U$ is obtained by solving the force balance in the swimming direction using MATLAB's \textit{fzero} function, which runs until a relative error of $10^{-16}$ is reached. A variety of shapes are provided as the initial guess for the starting of the optimization. The optimization calculation is iterated by taking the converged shape of the previous calculation as the initial guess until the shapes acquired in two successive calculations are consistent. The optimal shape obtained does not vary with the initial guess.

We validate our approach by solving the optimal shape for the Newtonian case. For Newtonian swimming, the swimming speed $U$ can be obtained by a simple matrix inversion due to linearity. The optimal shape obtained from our numerical approach agrees with the analytical solution of Lighthill \cite{lighthill1975mathematica}.

\subsection{Numerical solution to the equations of motion for finite swimmers}
\label{subsec:appendix_num}
The study of swimming characteristics requires solving the force and torque balance of the finite swimmer as formulated in Sec.~\ref{sec:form}. The force- and torque-free conditions posed in Eq.~(\ref{eq:Fbalance}) and (\ref{eq:Tbalance}) provide a system of non-linear ordinary differential equations (ODEs) for the swimmer's linear and angular velocities in terms of its instantaneous location and orientation. The instantaneous velocities in turn, once obtained,  can be integrated over time to determine the trajectory, location and orientation of the swimmer.

Having assumed that the centerline of the waveform is initially aligned with the $x-$axis of the lab frame, we solve the swimming problem numerically. Starting at $t=0$ with a time step $\Delta t$, we denote $t_i = i\Delta t$. With this notation, we employ a second order multi-step finite difference method to discretize the ODEs such that
\begin{align}
\bx_{i+1} =\frac{4}{3}\bx_{i}-\frac{1}{3}\bx_{i-1}+\frac{2\Delta t}{3}(2\dot{\bx}_i-\dot{\bx}_{i-1}).
\end{align}
We do similarly for $\theta_{i+1}$, and then $\Theta_{i+1}$ can be computed. To initialize this numerical scheme, we need both $[\bx_0, \theta_0]$ and $[\bx_1, \theta_1]$. At the first time step, $[\bx_1, \theta_1]$ is computed using the Runge-Kutta fourth order method. At each time step ($i \geq 1$), we obtain $[\dot{\bx}_i, \dot{\theta}_i]$ by solving the integral equations for force and torque balance using Gauss-Legendre quadrature integration coupled with MATLAB's \textit{fsolve} routine. We use $m$ points along the filament for the Gauss-Legendre integration method. The \textit{fsolve} routine in Matlab Optimization Toolbox attempts to solve a system of equations by minimizing the sum of squares of all the components. We set the termination tolerance on both the function value and independent variables to $10^{-14}$. We generally use $m=1000$ points along the filament and $T_{m}=500$ time steps for one period $T$ of the motion. The number of Fourier modes is taken as $n^{*}=100$. Further increasing of the number of spatial points or time steps have no significant influence on the accuracy of the results.  All the numerical simulations are performed using MATLAB.

\newpage

\bibliography{ref}
\end{document}